\documentclass[a4paper,twocolumn,
english,aps,pre,floatfix,superscriptaddress,showpacs]{revtex4-1}
\usepackage[T1]{fontenc}
\usepackage[latin1]{inputenc}
\usepackage{amsmath}
\usepackage{babel}
\usepackage{graphics}
\usepackage{amssymb}


\begin{document}
\title
{Scaling behavior of a square-lattice Ising model with competing interactions
in a uniform field}

\author {S. L. A. \surname{de Queiroz}}
\email{sldq@if.ufrj.br}
\affiliation{Rudolf Peierls Centre for Theoretical Physics, University of
Oxford, 1 Keble Road, Oxford OX1 3NP, United Kingdom}
\affiliation{Instituto de F\'\i sica, Universidade Federal do
Rio de Janeiro, Caixa Postal 68528, 21941-972
Rio de Janeiro RJ, Brazil}

\date{\today}

\begin{abstract}
Transfer-matrix methods, with the help of finite-size scaling and 
conformal invariance concepts, are used to investigate the critical behavior
of two-dimensional square-lattice Ising spin-$1/2$ systems with
first- and second-neighbor interactions, both antiferromagnetic,
in a uniform external field. On the critical curve separating 
collinearly-ordered and paramagnetic phases, our estimates of 
the conformal anomaly $c$ are very close to unity, indicating the
presence of continuously-varying exponents. This is confirmed by direct
calculations, which also lend support to a weak-universality picture;
however, small but consistent deviations from the Ising-like values
$\eta=1/4$, $\gamma/\nu=7/4$, $\beta/\nu=1/8$ are found.
For higher fields, on the line separating row-shifted $(2 \times 2)$
and disordered phases, we find values of the exponent $\eta$ very close to zero.
\end{abstract}
\maketitle
 
\section{Introduction} 
\label{intro}
The study of frustration in magnetism has been a very active
field of research in the recent past, both theoretically and
experimentally. While experimentally-realizable frustrated
magnets typically have a closer correspondence to quantum
(i.e., Heisenberg or $XY$) spin models than to classical, Ising-like
ones, their behavior turns out to be rather intricate. Thus,
theoretical and/or numerical investigation of frustrated classical spin
systems may, by virtue of their simplified character, help
unravel some basic features which are common to frustrated
magnets in general.

In this paper we investigate two-dimensional spin-$1/2$ Ising systems on a square
lattice with first- and second-neighbor couplings, both antiferromagnetic,
in the presence of a uniform magnetic field. The Hamiltonian is given by:
\begin{equation}
{\cal H}=J_1 \sum_{\rm NN}\sigma_{i}\,\sigma_{j}+
J_2 \sum_{\rm NNN}\sigma_{i}\,\sigma_{j} -H \sum_{i} \sigma_{i}\ ,
\label{eq:def}
\end{equation}
where $J_1$, $J_2 > 0$, NN and NNN stand respectively for next-neighbor and
next-nearest-neighbor pairs, and $\sigma_{i}=\pm 1$. Here, all fields,
coupling strengths and temperatures are given in units of $J_1$,
unless otherwise stated. We have kept $J_2=1$ in all calculations reported
in this work, except at the end of Sec.~\ref{sec:h=0}, where $J_2=2$ and $0.75$
were briefly considered (both for $H=0$).  

In line with the initial considerations given above, this may be considered
a classical approximation for the $J_1-J_2$ (Heisenberg) model~\cite{fm1,fm2}.
Nevertheless, as shown in the following, the model described by
Eq.~(\ref{eq:def}) exhibits intricate features of its own, several of
which are not fully understood so far. 
Depending on the relative strength
of the associated parameters, such setup of competing interactions
can generate various types of ordered phases at low temperature; more often than not,
the transitions between these and the high-temperature (paramagnetic) state 
do not belong to the standard Ising universality class~\cite{yl09,kalz08}.   

The problem studied here 
has been analyzed by several numerical techniques in the past;
Refs.~\onlinecite{yl09,kalz08} provide excellent summaries of earlier work, as well
as illustrations of the use of up-to-date Monte-Carlo (MC) simulation
techniques for this case. 

We use transfer-matrix (TM) methods~\cite{fs2}, in conjunction with 
finite-size scaling~\cite{barber}  and 
conformal invariance~\cite{cardy} concepts, to determine the location of 
the phase boundaries of systems described by Eq.~(\ref{eq:def}),
and the universality classes of the associated phase transitions. TM methods,
especially in the strip geometry used in this work,
are to some extent complementary to MC simulations, in that
they provide straightforward procedures for evaluation of the 
conformal anomaly, or central charge~\cite{bcn86},
as well as the decay-of-correlations exponent $\eta$ (via the amplitude-exponent 
relationship~\cite{cardy84}). Both quantities play an important role in the
identification of the universality classes pertaining to phase transitions
in two-dimensional systems, and neither is directly accessible via MC
methods (although direct estimates of $\eta$ can be produced by following
the decay of spin-spin correlations with distance in an MC context, no simple
relationship applies, such as the one given by conformal invariance 
on strips~\cite{cardy84}). On the other hand, similarly to MC techniques,
TM calculations also provide estimates of critical temperatures, specific heats, 
magnetizations, and susceptibilities. 

In Section~\ref{sec:calc} we recall the calculational methods used, 
as well as the finite-size scaling concepts and techniques employed in the analysis
of our results. Our numerical results for $H=0$ are given in 
Section~\ref{sec:h=0}, and those for $H \neq 0$ in 
Section~\ref{sec:h>0}.
Finally, in Sec.~\ref{sec:conc}, concluding remarks are made.

\section{Calculational Method and Finite-size scaling}
\label{sec:calc}

We set up the TM on strips of width $N$ sites, with periodic boundary conditions 
across. The coordinate axes coincide with the directions of the first-neighbor
bonds. We used $4 \leq N \leq 22$. For comparison, earlier TM studies of this 
problem~\cite{kkg83,akg84} could only reach $N=12$. For the case $J_2=1/4$
in zero external field, results for $N \leq 18$ are available~\cite{nb98}.

With $\lambda_1$, $\lambda_2$ being the two largest eigenvalues (in absolute value) 
of the TM, the dimensionless free energy per site is given by 
$f_N(T)=N^{-1}\ln \lambda_1$, while 
$\kappa_N(T)=\ln |\,\lambda_1 /\lambda_2\,|$
is the inverse correlation length on a strip of width $N$ sites~\cite{fs2}.

It must be stressed that we do not make any assumptions about
symmetry properties of the TM's eigenvectors. Starting from the full
set of $2^N$ basis vectors, the eigenvector $|\,v_1 \rangle$,
corresponding to $\lambda_1$, is isolated by the power method, while
$|\,v_2 \rangle$ is again extracted via the power method, combined with
repeated Gram-Schmidt orthogonalization to $|\,v_1 \rangle$. This way we  
make sure that the most strongly diverging correlation length is evaluated,
that is, the one which truly corresponds to the order parameter for the
transition under scrutiny. This is especially relevant in the present case,
where critical lines corresponding to order parameters of differing
symmetries can become very close (see Section~\ref{sec:h>0} below). 

Here we assume that the transition is always of second order, which is
implicit in the statements made in the preceding paragraph. By now this seems
out of doubt, at least for $J_2 \geq 1$ which is the parameter range of interest 
here~\cite{yl09,kalz08}. 
For fixed $J_2$ and $H$, say, we locate the approximate 
($N$-dependent) critical temperature $T^\ast_N$ by solving the basic equation 
of the phenomenological renormalization group (PRG)~\cite{fs2}:
\begin{equation}
N \kappa_N(T)=N^{\prime} \kappa_{N^{\prime}}(T)\ .
\label{eq:prg}
\end{equation}
Depending on the shape of the critical curve, it may be more convenient
to keep $T$ fixed and vary $H$, in which case the critical behavior
is expressed in terms of $|H-H_c|$, and Eq.~(\ref{eq:prg}) gives $H^\ast_N$.
The strip widths $N$ and $N^{\prime}$ 
are to be taken as close as possible for improved convergence of results 
against increasing $N$. In order to obey ground-state symmetries,
here we used only even $N$, $N^\prime$, so $N^{\prime}=N-2$. 
For $H=0$ and $J_2<1/2$, in which case the ordered phase is 
N\'eel-like~\cite{kalz08}, one can use both odd and even $N$; indeed,
good results are found from PRG with $N^\prime=N-1$ in that region~\cite{nb98}. 
For $J_2>1/2$, we found that: (i) the latter procedure does not give physically 
meaningful solutions for Eq.~(\ref{eq:prg}); and (ii) although PRG with both $N$ 
and $N^\prime$ odd gives the same
limiting $T_c$ with $N \to \infty$ as when both strip widths are even (albeit
with much slower convergence), estimates of quantities other than the critical
temperature are unreliable.

Estimates of the thermal exponent $y_T=1/\nu$ are given by~\cite{fs2}:
\begin{equation}
y_T=1+\frac{\ln (\kappa_N^\prime/\kappa_{N^\prime}^\prime)}{\ln (N/N^\prime)}\ ,
\label{eq:y_prg}
\end{equation}
where $\kappa_N^\prime$, $\kappa_{N^\prime}^\prime$ are temperature 
derivatives of the inverse correlation lengths, taken at $T^\ast_N$.
Finite-$N$ estimates of the exponent $\eta$ are given by the
conformal invariance relation~\cite{cardy84}:
\begin{equation}
\eta_N=\pi^{-1}\,N\kappa_N(T^\ast_N)\ .
\label{eq:eta_prg}
\end{equation}
The convergence of finite-$N$ approximants given by Eqs.~(\ref{eq:prg}),
(\ref{eq:y_prg}), and~(\ref{eq:eta_prg}) towards their $N \to \infty$
values has been extensively discussed~\cite{dds82,nb82,pf83,bg85,bg85b}.
For (unfrustrated) Ising-like systems on strips with periodic
boundary conditions across, the rate of convergence goes like 
\begin{equation}
X_N-X_\infty =a\,N^{-\omega}\ ,
\label{eq:prg_conv}
\end{equation} 
with $\omega \approx 3$ for $X=T^\ast$, and $\omega \approx 2$ for $X=y_T$
(for some simple cases, this can be shown analytically~\cite{dds82,bg85,bg85b}).
By taking sets of three successive finite-$N$ estimates, one can use $\omega$
as an adjustable parameter in Eq.~(\ref{eq:prg_conv}), 
and produce a new, shorter, sequence which
can then be iterated again, and so on. Such iterated three-point fit  
technique can produce very accurate final estimates of critical 
quantities~\cite{nb82,bn82,bn85}. 

Once $T_c$ is found, as described above, to good accuracy (or if its exact
value is known, for example via duality arguments~\cite{bn82}), sequences
of assorted quantities can be evaluated at the extrapolated 
critical point, for increasing $N$. 
From these, one can usually extract estimates
of critical exponents which converge faster and more smoothly
than if the calculations were done at the respective 
pseudo-critical temperatures $T^\ast_N$~\cite{dds82,nb83}.
One is interested in (per site) specific heats, susceptibilities, and magnetizations,
which behave as~\cite{barber}:
\begin{eqnarray}
C_N(T_c)=C_0+a_CN^{\alpha/\nu}\ ;\nonumber \\
\chi_N(T_c) =a_\chi N^{\gamma/\nu}\ ;\nonumber \\
m_N(T_c)=a_m N^{-\beta/\nu}\ .
\label{eq:fss2}
\end{eqnarray}  
Both $C_N$ and $\chi_N$ are found from suitable second derivatives of the free 
energy~\cite{bn82}. The exponent ratio $\alpha/\nu$ can then
be extracted from three-point fits of sequences of $C_N$, as explained in 
connection with Eq.~(\ref{eq:prg_conv}). For $\gamma/\nu$, one initially
obtains a sequence of exponent estimates via two-point fits of susceptibility
data, and then proceeds to extrapolating such a sequence via three-point
fits~\cite{bn82}.

The spontaneous magnetization $m_N$ is difficult to calculate in
a finite-size scaling context, because it is identically zero for a finite
system.
For quantum chains at $T=0$ this problem can be overcome~\cite{hamer82},
by exploiting the fact that there the largest eigenvalue of the TM gives 
the internal energy: in a first-order degenerate perturbation scheme, 
appropriate consideration of non-diagonal matrix elements enables
one to extract the magnetization in the zero-field limit.
For classical spins on strips, the corresponding eigenvalue of the TM 
gives the free energy instead, and the perturbation-theory procedure used
for quantum systems~\cite{hamer82} cannot be translated to our case.

In this work we estimated the finite-size magnetization exponent $\beta/\nu$ by
calculating the average squared magnetization per column 
at $T_c$, $\langle M^2\rangle$. Considering, for example, a ferromagnet, denoting by 
$\sigma\equiv \{\sigma_1\,\cdots\,\sigma_N\}$ the $2^N$ column basis states,
and with ${\tilde \psi} (\sigma)$,
$\psi (\sigma)$ being respectively the dominant left and right eigenvectors of the 
TM, one has~\cite{fs2}:
\begin{equation}
\langle M^2\rangle =\frac{\sum_{\sigma}{\tilde \psi}(\sigma)\left(\sum_{i=1}^N 
\sigma_i\right)^2\,\psi (\sigma)}{\sum_{\sigma}{\tilde \psi}(\sigma)
\,\psi (\sigma)}\ .
\label{eq:magsq}
\end{equation}
At the critical point, one should have:
\begin{equation}
\frac{1}{N}\,\langle M^2\rangle^{1/2} \sim N^{-\beta/\nu}\ .
\label{eq:fssbeta}
\end{equation}   
For a square-lattice antiferromagnet with only first-neighbor interactions,
$\sigma_i$ in Eq.~(\ref{eq:magsq}) must be replaced by $(-1)^i\,\sigma_i$,
so the staggered character of the order parameter is properly taken into
account. The corresponding adaptation for the system of interest
here is discussed in Section~\ref{sec:nr} below.
We tested this procedure, with the appropriate (uniform or staggered)
version of the magnetization, on the following square-lattice Ising systems: 
(i) ferromagnet with nearest-neighbor couplings only; (ii) ferromagnet with
first- and second-neighbor interactions, $J_2=1$; and (iii) antiferromagnet with
first- and second-neighbor interactions, $J_2=1/4$ (for which the
ordered state is N\'eel-like~\cite{nb98}). In all three cases, $T_c$ is
known either exactly or to a very good approximation, and the
transition is in the Ising universality class~\cite{nb98}, so $\beta/\nu=1/8$. 
In order to gauge the likely systematic errors for our intended final application
(see Section~\ref{sec:nr}), 
we considered $4 \leq N \leq 22$, and only
even $N$. All resulting sequences gave estimates of $\beta/\nu$ monotonically growing
with $N$, pointing to extrapolated values between $0.1240$ and $0.1247$, so the
systematic error is less than $1\%$ for this range of $N$.

Another quantity of interest to be calculated at $T_c$ is the conformal anomaly
$c$, given by the $N^{-2}$ finite-size correction of the critical free 
energy per site~\cite{bcn86}. For the present case of strips with periodic boundary
conditions across, one has:
\begin{equation}
f_N(T_c)=f_0+\frac{\pi\,c}{6N^2}+{\cal O}(N^{-4})\ .
\label{eq:c}
\end{equation}
While models with $c<1$ are associated with universality classes with fixed
values of the critical exponents, those with $c \geq 1$ can have 
continuously varying exponents~\cite{cardy87,dif97}. As shown below, there are
strong indications that the model studied here belongs to the latter
category (for $H=0$, this has already been pointed out in  Ref.~\onlinecite{kalz11}).
   
Additionally, one can both double-check the robustness of extrapolations 
of $T_c$ and $y_T$ 
from Eqs.~(\ref{eq:prg}) and~(\ref{eq:y_prg}), as well as investigate the other
quantities of interest, by scanning the neighborhood of the critical point with the
help of finite-size scaling ideas~\cite{barber}. Taking, for instance,
$\eta_N(T) \equiv \pi^{-1}\,N \kappa_N(T)$, and allowing for corrections to
scaling, we write~\cite{has08,dq09}:
\begin{equation}
\eta_N(T)=f(u)+ N^{-\omega} g(u)\ ,\quad u \equiv N^{1/\nu}(T-T_c)\ .
\label{eq:fss}
\end{equation}
where $\omega >0$ is the exponent associated with the leading irrelevant 
operator [$\,$see Eq.~(\ref{eq:prg_conv})$\,$]. 
Close enough to $T_c$ the scaling functions 
in Eq.~(\ref{eq:fss}) should be amenable to Taylor expansions. One has:
\begin{equation}
\eta_N(T)=\eta+ \sum_{j=1}^{j_m} a_j\,u^j+
N^{-\omega}  \sum_{k=0}^{k_m} b_k u^k\ ,
\label{eq:fss2b}
\end{equation}
where $\eta$ is to be compared with the $N \to \infty$ extrapolated value
of the $\eta_N$ of Eq.~(\ref{eq:eta_prg}). 

One looks for values of $T_c$, $\nu$, $\omega$ and the $\{a_j,b_k\}$
which optimize data collapse upon plotting $\eta_N(T)-N^{-\omega}g(u)$
against $u$. In practice, good fits are generally found with $j_m$, $k_m$
not exceeding $2$ or $3$~\cite{has08,dq09}.  

Considering now the finite-size susceptibility $\chi_N(T)$,
finite-size scaling~\cite{barber} suggests a form
\begin{equation}
\chi_N(T) = N^{\gamma/\nu}\,f_\chi\,(u)\ ,
\label{eq:chisc}
\end{equation}
where $\gamma$ is the susceptibility exponent. 
Following Refs.~\onlinecite{has08,dq09}, we write (again, allowing for
corrections to scaling):
\begin{equation}
\ln \chi_N(T) =\frac{\gamma}{\nu} \ln N +\sum_{j=1}^{j_m} a_j^{\,\chi}\,u^j
+ N^{-\omega}  \sum_{k=0}^{k_m} b_k^{\,\chi}\,u^k\ .
\label{eq:fss3}
\end{equation}
In order to reduce the number of fitting parameters, it is usual to keep $1/\nu$
and $T_c$ fixed at their central estimates obtained, e.g., via Eqs.~(\ref{eq:fss}) 
and~(\ref{eq:fss2b}), allowing $\gamma/\nu$ to vary. 

Expressions similar to Eq.~(\ref{eq:fss3}) can be written for magnetizations
and specific heats, yielding estimates of the exponents $\beta/\nu$ and $\alpha/\nu$. 

In the present context, one should interpret the 
exponent $\omega$ in  Eqs.~(\ref{eq:prg_conv}), (\ref{eq:fss}), and~(\ref{eq:fss3}) 
as an effective one, representing all orders of corrections to scaling (which may
also turn out to have rather different amplitudes for different quantities).
Thus, in practice a somewhat broad range of results (say,
$1 \lesssim \omega \lesssim 3$) can be accepted when considering data
collapse optimization for distinct quantities related to the same problem.

\section{Numerical Results}
\label{sec:nr}

\subsection{$H=0$}
\label{sec:h=0}

In zero external field, for $J_2 <1/2$ the ground-state ordering
is of the N\'eel type, with the two sublattices aligned antiparallel
to each other; the transition is second-order, in the Ising universality
class~\cite{nb98}. At $J_2=1/2$ the ground state is macroscopically degenerate, and 
the critical temperature is zero~\cite{kalz08}.  
For $J_2>1/2$ the lowest energy corresponds to collinear
order, with alternating rows (or columns) of parallel spins. For $J_2 \gtrsim 1/2$
the transition is first order, and evidence has been found that it 
remains so, at least up to 
$J_2 \approx 0.9$~\cite{kalz08,kalz11}. As $J_2$ increases further, the second-order
character returns. The bulk of extant evidence~\cite{yl09,kalz11,mkt06,mk07,kim10}
indicates that the transition is second order for $J_2=1$.
Finally, for $J_2 \gg 1$ one has a picture of two weakly-coupled
antiferromagnetic lattices, thus in this limit $T_c/J_2$ approaches the
Ising value, $2/\ln (1 +\sqrt{2})$.

For $J_2=1$, recent estimates of the critical temperature are:
$T_c=2.0823(17)$~\cite{mkt06}; $2.0838(5)$\cite{mk07}; $2.0820(4)$~\cite{yl09};
and $2.0839(12)$~\cite{kim10}. By solving Eq.~(\ref{eq:prg}), we found
a well-behaved sequence of $T^\ast_N$ values, extrapolating to $T_c=2.08195(5)$
via three-point fits, with $\omega \approx 3$. The sequences 
for $y_T$ and $\eta$ from Eqs.~(\ref{eq:y_prg}) and~(\ref{eq:eta_prg}) 
extrapolate respectively to $y_T=1.188(2)$ and $\eta=0.2341(1)$,
again via three-point fits.

Evaluating $\eta_N(T)$ in the region around $T_c$ , and employing 
Eqs.~(\ref{eq:fss}) and~(\ref{eq:fss2b}), gave $T_c=2.08197(5)$;  
$y_T=1.182(3)$, and $\eta=0.2342(1)$, with $\omega \approx 1.9$.
We used $j_m=2$, $k_m=1$ in Eq.~(\ref{eq:fss2b}).
Thus there is a satisfactory degree of consistency between the two
methods of evaluation of critical quantities.

Our numerical value for $\nu=y_T^{-1}=0.844(4)$ [$\,$from averaging
over the two results above$\,$] is to be compared to 
$\nu=0.8292(24)$~\cite{mkt06}; $0.8481(2)$\cite{mk07}; $0.847(4)$~\cite{yl09};
and $0.847(1)$~\cite{kim10}. The value $\eta=0.20(1)$ was found
by direct MC evaluation of critical correlation functions 
on $N \times N$ geometries~\cite{kalz11}.      

By evaluating quantities at the extrapolated $T_c$, we found $\eta=0.23415(5)$;
as expected, this is even more accurate than extrapolating the sequence
of finite-$N$ values estimated at the respective fixed points $T^\ast_N$.

For calculation of the zero-field susceptibility, the specific properties of
the collinear order parameter were taken into account as follows. 
For a fixed coordinate direction, say $x$, along which the TM proceeds,
the critical wavevector is degenerate, being either $(\pi/a)\,{\hat x}$ 
or $(\pi/a)\,{\hat y}$, with $a$ being the lattice parameter. 
One thus has to take both (equally probable
and mutually exclusive) possibilities into account and 
average the partial contributions given by each.
As might be expected, we found both contributions to be of similar amplitudes
(within $\approx 10\%$ of each other for fixed $N$); separate fits of each to
power-law forms gave apparent exponents differing by less
than $1\%$. The latter discrepancy can be ascribed to residual lattice
effects, and is expected to vanish for larger $N$, out of reach of our
TM implementations at present. 
  
Our final result was $\gamma/\nu=1.772(1)$.
Estimating $\gamma/\nu$ via Eq.~(\ref{eq:fss3}) gave
$\gamma/\nu=1.775(1)$, slightly higher than the previous estimate but within
three (rather narrow) error bars. Since the uncertainties quoted 
refer exclusively to the fitting procedures, i.e., no account is taken of likely 
systematic errors, one might err on the side of caution and allow for
somewhat large uncertainties. Averaging over the two values found, we quote 
$\gamma/\nu=1.773(4)$. This way our results might be considered marginally 
compatible with $\gamma/\nu=1.750(12)$, quoted in Ref.~\onlinecite{yl09},
although it seems much harder to stretch our error bars to include the
value $7/4$, which would be consistent with a weak-universality picture
of $\gamma/\nu$, $\beta/\nu$, $(2-\alpha)/\nu$ 
sticking to the respective Ising values~\cite{s74}.
Ref.~\onlinecite{mkt06} quotes the range $1.71-1.79$ for $\gamma/\nu$, based 
on three diferent fitting methods.

For the calculation of $\beta/\nu$, we used as column magnetization
in Eq.~(\ref{eq:magsq}) the following quantity:
\begin{equation}
\langle M^2 \rangle =\frac{1}{2}\left[ \langle M_{\rm u}^2\rangle +
\langle M_{\rm st}^2\rangle\,\right]\ ,
\label{eq:m_comb}
\end{equation}
where $M_{\rm u}$, $M_{\rm st}$ are respectively uniform and staggered column 
magnetization. This choice reflects the collinear nature of the ground state,
with its orientational degeneracy, and closely corresponds to the order
parameter used in the MC simulations of Ref.~\onlinecite{yl09}. See
the arguments invoked above for the susceptibility calculation.    
Similarly to the test cases described in 
Section~\ref{sec:calc}, we found exponent estimates monotonically growing with
$N$; the extrapolated result is $\beta/\nu=0.121(2)$, where an {\em ad hoc}
doubling of the uncertainty found from fits has been incorporated, in order to 
allow for the small bias shown in tests.   

We also evaluated critical specific heats.
Finite-size specific-heat sequences can prove unwieldy to extrapolate, even when
the exponent $\alpha$ is positive, as in the case of the two-dimensional 
three-state Potts ferromagnet~\cite{bn82}. Here, three-point fits of $N$,
$N-2$, $N-4$ data gave $\alpha/\nu$ increasing from $\approx 0.31$ ($N=10$) 
to $\approx 0.33$ ($N=20$), albeit with small oscillations; a quadratic fit
of such values against $1/N$ then gave $\alpha/\nu=0.351(12)$. 

The above results are to be compared to $\beta/\nu=0.122(4)$, $\alpha/\nu=0.357(8)$,
both from Ref.~\onlinecite{yl09}, and $\alpha/\nu=0.412(5)$~\cite{mkt06}. 
Recalling the Rushbrooke scaling relation, $\alpha+2\beta+\gamma=2$,
our estimates give $\alpha/\nu+2\beta/\nu+\gamma/\nu=2.366(13)$,
while $2/\nu=2.370(10)$. Similarly, one has $(2-\alpha)/\nu=2.020(18)$.
Given the rather large uncertainties found in the analysis of specific heat behavior,  
we do not believe that any actual breakdown of hyperscaling is present.  

We calculated free energies at $T_c$, and fitted them to a quadratic form in
$1/N^2$, thus extracting estimates of the conformal anomaly~\cite{bcn86}.
From fits of data in the range $[N_{\rm min},22]$, with
$4 \leq N_{\rm min} \leq 14$, we found estimates of $c$ decreasing monotonically from
$1.074(1)$ for $N_{\rm min}=4$ to $1.056(1)$ for $N_{\rm min}=14$. Uncertainties
quoted relate exclusively to the fitting procedure. This range of
estimates compares favorably with the corresponding result 
from Ref.~\onlinecite{kalz11}, $c=1.0613(6)$. It must be kept
in mind that what one is seeing most likely amounts to 
strong crossover effects distorting a picture where $c=1$~\cite{kalz11}.

We also considered $J_2=2$, for which case we obtained, from 
extrapolating sequences generated via Eqs.~(\ref{eq:prg}), (\ref{eq:y_prg}),
and~(\ref{eq:eta_prg}) , $T_c/J_2=2.2248(1)$, $y_T=1.052(1)$,
$\eta=0.2391(3)$. Recent results are: $T_c/J_2=2.226(5)$~\cite{mk07}, 
and $2.227(5)$~\cite{kim10}; $y_T=1.066(1)$~\cite{kim10}.
Evaluation of finite-size susceptibilities, magnetizations,
and specific heats at $T_c$ gave $\gamma/\nu=1.756(2)$; $\beta/\nu=0.120(2)$;
$\alpha/\nu \lesssim 0.1$ (estimates for the latter quantity were plagued
by the same sort of irregularities reported for $J_2=1$ above). 
Overall, these values 
are consistent with a picture of continuously-varying
exponents, approaching the Ising ones as $J_2$ increases~\cite{yl09,kalz11}. 
Conformal-anomaly estimates are very close to $c=1.010$; again, this is
consistent with the trend towards $c=1$, followed by fitted results
upon increasing $J_2$, found in Ref.~\onlinecite{kalz11}.

Finally, we made $J_2=0.75$, which is expected to correspond to a first-order
transition~\cite{kalz11}, thus in principle the ideas behind
Eq.~(\ref{eq:prg}) do not apply. 
Indeed, instead of varying monotonically with increasing $N$,
the solutions of Eq.~(\ref{eq:prg}) initially went up, to $T_c \approx 1.432$ at
$N=12$, and then became approximately constant for larger $N$; the $\eta$ estimate
from Eq.~(\ref{eq:eta_prg}) also initially increased, up to $\approx 0.243$ at
$N=14$ and $16$, then started decreasing for larger $N$. This indicates
a correlation length which at the very least grows slower than $N$, and
possibly saturates at scales which are out of reach of our TM calculations,
that is, a weakly first-order transition~\cite{kalz08,kalz11}. Notwithstanding 
the lack of conceptual justification for using Eq.~(\ref{eq:prg}), it should be 
noted  that $T_c \approx 1.43$ is in rather good agreement with MC 
estimates (see Figure 3 of Ref.~\onlinecite{kalz08}).

\subsection{$H \neq 0$}
\label{sec:h>0}

For $J_2=1$, the ground state is still collinear for $H<4$, whereas
for $4<H<8$ it becomes a row-shifted $(2\times 2)$ state~\cite{yl09}.
The latter consists of alternating ferro- and antiferromagnetically
ordered rows (or columns), with the ferromagnetic ones parallel
to the field. The added degree of freedom (relative
to a $2 \times 2$ state) is that the antiferromagnetic chains can slide
freely relative to each other, at zero energy cost.  
At $H=4$ and $8$, $T_c=0$ because of macroscopic ground-state degeneracy.
The maximum critical temperature for the transition between row-shifted
$(2 \times 2)$ order and the paramagnetic phase has been estimated as 
$\approx 0.73$, at $H \approx 6$~\cite{yl09}.

In order to make contact with previous results, we initially considered 
two points on the collinear-paramagnetic transition line, 
respectively at $H=2.5$ and $3.3$~\cite{yl09}. 

For $H=2.5$, from  extrapolating sequences generated via 
Eqs.~(\ref{eq:prg}), (\ref{eq:y_prg}), and~(\ref{eq:eta_prg}),
we found $T_c=1.6846(1)$,  $\eta=0.2335(1)$. 
A noticeable trend reversal was observed for $y_T$; after decreasing
from  $1.08$ to $1.072$ between $N=8$ and $12$, it starts increasing smoothly, 
reaching $1.075$ at $N=22$. Extrapolating $N \geq 12$ data, we
found $y_T=1.090(4)$. Evaluating susceptibilities
at the extrapolated $T_c$ resulted in $\gamma/\nu=1.779(4)$, and analysis
of magnetizations gave $\beta/\nu=0.122(1)$. 
Ref.~\onlinecite{yl09} gives:
$T_c=1.6852(3)$; $y_T=1.056(8)$; $\gamma/\nu=1.750(14)$, $\beta/\nu=0.118(3)$.
Finally, the conformal anomaly was estimated as $c=1.066(1)$.

Following the same procedure as above, we obtained for $H=3.3$:
$T_c=1.3331(5)$; $y_T=0.940(3)$ (this time with no trend reversal upon increasing $N$);
$\eta=0.2338(3)$.  Finite-size susceptibility scaling at $T_c$
gave $\gamma/\nu=1.781(5)$, and magnetizations, $\beta/\nu=0.135(2)$.
The evolution of $\beta/\nu$ along the collinear-paramagnetic
phase boundary is analyzed towards the end of this Section.
Ref.~\onlinecite{yl09} quotes: $T_c=1.3335(6)$; $y_T=0.907(7)$; 
$\gamma/\nu=1.751(14)$, $\beta/\nu=0.130(5)$. Our estimate for the 
conformal anomaly is $c=1.042(1)$.

The above results show very good numerical agreement with existing ones as
regards critical temperatures; also, a picture of continuously-varying
exponents such as $\nu$ and $\gamma$ is confirmed, both directly
and from the conformal anomaly results, which are very close to unity. 
Our values for $\gamma/\nu$ and $\eta$ do seem consistent with a weak-universality
scenario along the collinear-paramagnetic phase boundary; however,
they indicate small but consistent deviations from the Ising-like picture of 
$\gamma/\nu=7/4$, $\eta=1/4$.

In order to investigate the latter point in detail, we proceeded to
evaluating $\eta$ and $c$ along the full extent of the phase boundary.

We first extrapolated, to $N\to \infty$, the $(T,H)$ values obtained for sequences of
solutions of Eq.~(\ref{eq:prg}), with increasing $N$ and either
$H$ fixed, or (typically for lower temperatures) $T$ fixed.
This was done both for the collinear-paramagnetic critical line and for
that separating the row-shifted $(2 \times 2)$ and paramagnetic phases.
In the former case, we fitted finite-$N$ data with $N=14-20$ to the single-power
form, Eq.~(\ref{eq:prg_conv}), finding good convergence with 
$3 \lesssim \omega \lesssim 4$ everywhere on the phase boundary.
For the latter, we had extremely slow convergence of our PRG calculations, which
limited us in practice to $N \leq 16$. Probably (at least partially) as
a consequence of this, the single-power form
produced rather low  adjusted values of $\omega$, in the neighborhood of
$0.5$, which is usually interpreted as indicating strong corrections to 
scaling. We thus resorted to {\em ad hoc} parabolic fits in $N^{-2}$,
adjusting $N=10-16$ data to this latter form. 

The behavior in the low-temperature region near $H=4$, where the two
distinct phase boundaries become close to each other, 
is of special interest since it has been suggested that
an $XY$-like region might be present there~\cite{yl09,bl80}.
\begin{figure}
{\centering \resizebox*{3.3in}{!}{\includegraphics*{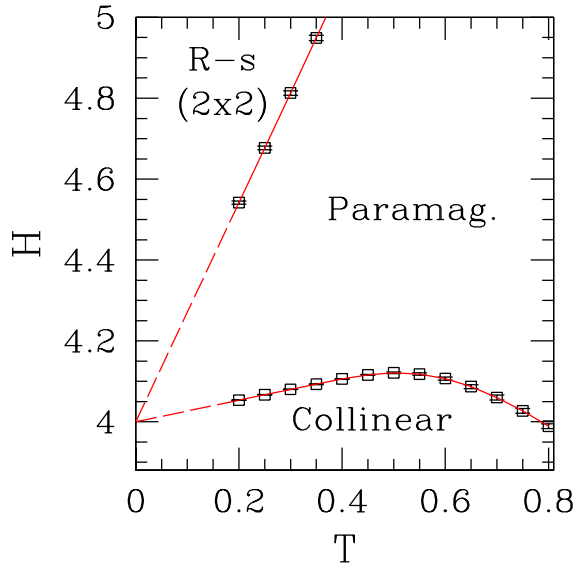}}}
\caption{(Color online) Phase diagram near $H=4$, showing phase boundaries:
collinear-paramagnetic, with reentrant behavior, and row-shifted [R-s] 
$(2 \times 2)$-paramagnetic. The points
are the results of extrapolating sequences of solutions of 
Eq.~(\protect{\ref{eq:prg}}), obtained with fixed $T$ and variable $H$ (see text). 
Uncertainties are smaller than symbol sizes. 
Each of the dashed lines at $T<0.2$ is the continuation of the 
best-fitting straight line joining points at $0.2 \leq T \leq 0.4$, 
on the respective phase boundary. 
} 
\label{fig:reentpd}
\end{figure}

In Figure~\ref{fig:reentpd} we show our results for the low-temperature part of the 
phase diagram, near $H=4$. Although numerical convergence difficulties 
prevented us from reaching $T<0.2$ for the largest strip widths, 
we managed to evince clearly-defined trends 
followed along both critical lines, on their approach to $T=0$. Below $T=0.4$,
both the collinear-paramagnetic and row-shifted $(2\times 2)$-paramagnetic
boundaries are, to a very good approximation, straight, and pointing
towards $(T,H)=(0,4)$ with respective slopes $0.267(1)$ and $2.713(5)$.
Concurring with Ref.~\onlinecite{yl09}, it appears very unlikely that an 
$XY$-like region, or a bicritical point at $T>0$, is present. Instead, 
all indications from our results are consistent with both critical
lines joining at a single bicritical point at $T=0$. 
Additionally, we found the maximum of the reentrance on the 
collinear-paramagnetic phase boundary to be $H =4.121(2)$ at $T=0.5$.
At $T=0.7$ we estimate $H=4.060(6)$. These are in rather
good agreement with the respective values $H=4.07(2)$ and $4.052(7)$,
quoted in Ref.~\onlinecite{yl09}.

Near $H=0$, the phase boundary has the expected parabolic shape~\cite{k87,dq09b}:
\begin{equation}
T_c(H)=T_c(0)-a\,H^2\ \ \quad (H \to 0)\ ,
\label{eq:lowh}
\end{equation} 
where we found $a=0.0595(3)$ by fitting data corresponding to $0 \leq H \leq 0.4$.

\begin{figure}
{\centering \resizebox*{3.3in}{!}{\includegraphics*{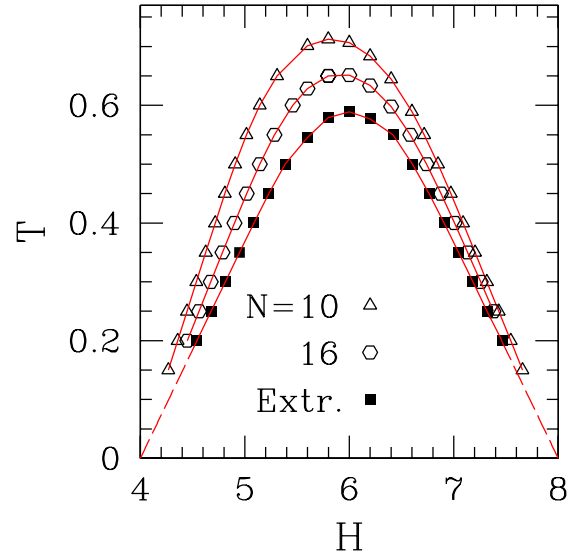}}}
\caption{(Color online) Phase diagram for high fields $4 \leq H \leq 8$, showing 
row-shifted [R-s]  $(2 \times 2)$-paramagnetic phase boundary: finite-$N$
solutions of Eq.~(\protect{\ref{eq:prg}}) for $N=10$ and $16$, and results
of extrapolation of $N=10-16$ curves (see text).
Uncertainties in the latter are smaller than symbol sizes. 
Each of the dashed lines at $T<0.2$ is the continuation of the 
best-fitting straight line joining points at $0.2 \leq T \leq 0.4$, 
on the respective section of the extrapolated phase boundary. 
} 
\label{fig:hfpd}
\end{figure}

As noted previously in Ref.~\onlinecite{akg84} and above, we generally found finite-size
effects to be much larger for the high-field ($4 \leq\ H \leq 8$) part of the
phase diagram. This is illustrated in Figure~\ref{fig:hfpd}, where finite-$N$ curves
with the solutions of  Eq.~(\ref{eq:prg}) for $N=10$ and $16$ are displayed
jointly with our final extrapolation (as described above).
At variance
with Ref.~\onlinecite{yl09}, where $T_c=0.7293(7)$ is reported at $H=6$,
our extrapolated value is $T_c(H=6)=0.589(4)$. Although alternative
procedures to our {\it ad hoc} parabolic extrapolations against $N^{-2}$
can certainly be devised, it must be noted that 
for this same field intensity the solution of Eq.~(\ref{eq:prg}) is
 $T^\ast_N =0.706$ already for $N=10$, and decreases systematically
with increasing $N$.     

At $(T,H)=(0,8)$, the initial slope $S=(dH_c/dT)_{T=0}$ of the critical curve 
gives an estimate of
the reduced critical chemical potential $\mu/k_BT_c$ for the hard-square 
lattice gas with first- and second-neighbor exclusion, via 
$\mu/k_BT_c=-2S$~\cite{akg84}. Our result for the chemical potential is $5.42$, 
to be compared with $4.70$~\cite{akg84}, and $4.91$~\cite{slotte}. This may
indicate that
our extrapolation procedures slightly underestimate the extent of the
row-shifted $(2 \times 2)$ phase.

\begin{figure}
{\centering \resizebox*{3.3in}{!}{\includegraphics*{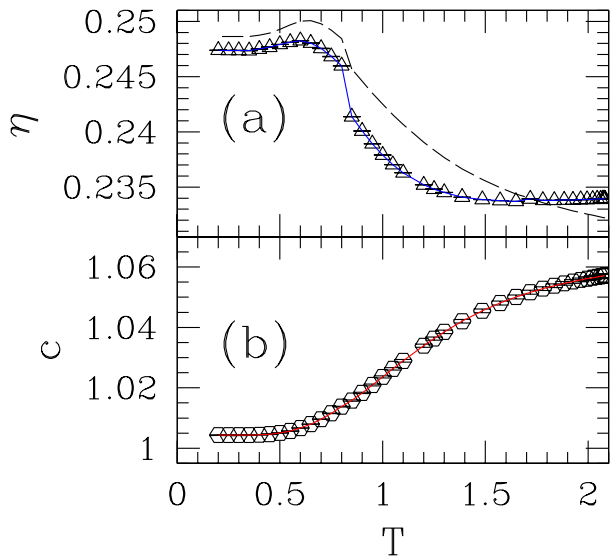}}}
\caption{(Color online) (a) Decay-of-correlations exponent $\eta$ and (b)
conformal anomaly $c$, calculated along the extrapolated collinear-paramagnetic
phase boundary. In (a), the dashed line gives estimates of $\eta$ from
strips of width $N=10$ sites, while points are extrapolations from sets
of finite-$N$ estimates, $N=14-20$. In (b), values of $c$ are estimated from
quadratic fits of free-energy data against $N^{-2}$ for $N=14-20$.
Uncertainties are smaller than symbol sizes. 
} 
\label{fig:eta_cc}
\end{figure}

Figure~\ref{fig:eta_cc} shows our results for $\eta$ and $c$  along the
extrapolated  location of the collinear-paramagnetic border, parametrized by $T$.
These are obtained from Eqs.~(\ref{eq:eta_prg}) and~(\ref{eq:c}), respectively.  
In part (a), comparison between $N=10$ estimates
and the final $N \to \infty$ extrapolation illustrates that residual
finite-size effects contribute towards overestimating the exponent $\eta$,
for all $0 < T \lesssim 1.78$ (the approximate point where all finite-$N$ curves cross).
On the other hand, for higher $T$ the finite-size corrections change sign.
The extrapolated $\eta \times T$ curve is to a large extent horizontal, 
near both the low- and
high-temperature ends of the phase boundary. We estimate $\eta=0.2476(3)$ for 
$T \leq 0.5$, and $\eta=0.2342(3)$ for $T \geq 1.3$. In the intermediate region
there is a crossover which becomes rather sharp around $T=0.8$, at the
upper end of the reentrant part of the phase diagram.
The conformal-anomaly estimates in part (b) show the same behavior found for
$H=0$ in Section~\ref{sec:h=0}, and in Ref.~\onlinecite{kalz11}, in that
they are always slightly above unity. Similarly to the $H=0$ case,
we also found that, upon fitting free-energy data in the range $[N_{\rm min},20]$, 
the estimates of $c$ always decrease upon increasing $N_{\rm min}$. 
Thus, an interpretation of the present results as consistent with $c=1$, albeit
affected by strong crossover effects, seems credible.

\begin{figure}
{\centering \resizebox*{3.3in}{!}{\includegraphics*{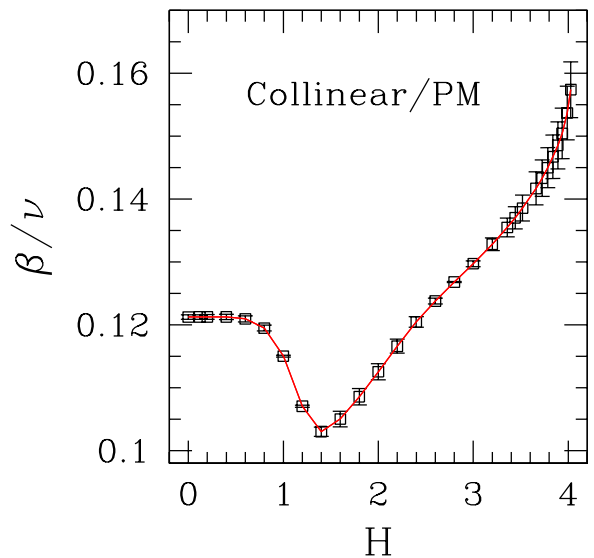}}}
\caption{(Color online) Finite-size magnetization exponent $\beta/\nu$
along collinear-paramagnetic (PM) phase boundary. The range of fields, 
$0 \leq H \leq 4.027$, on the horizontal axis corresponds, respectively,
 to $2.0820 \geq T \geq 0.75$ (see text). Each point is the result
of fitting finite-size data in the range $12 \leq N \leq 20$ to a
single-power law, $\langle M^2\rangle^{1/2} \sim N^{1-\beta/\nu}$.
See Eqs.~(\protect{\ref{eq:fssbeta}}) and~(\protect{\ref{eq:m_comb}}). 
} 
\label{fig:beta}
\end{figure}

We evaluated critical magnetizations, as given by Eq.~(\ref{eq:m_comb}),
along the collinear-paramagnetic phase boundary. Our calculations
did not converge for $T <0.75$, which approximately coincides with the reentrant
region. Thus, for the part of the critical boundary where we managed
to produce estimates of $\beta/\nu$, there is a one-to-one
correspondence between field and temperature.  
Our results are shown in Figure~\ref{fig:beta}, parametrized by $H$.
This way, it is easier to follow the evolution of quantities for low
fields than if we used $T$ for the horizontal axis, because
of the parabolic shape assumed by the critical curve in that region. 
One sees that the quality of fits generally deteriorates as $H$
increases; the shallow dip around $H \approx 1.5$ is possibly related
to slight inaccuracies in the determination of the extrapolated critical
line in that region. A more persistent trend is that towards increasing
values for larger $H$. We interpret this as signalling the onset
of the physical effects which give rise to reentrant behavior for even
larger fields. Indeed, a plausible explanation for the reentrance is, 
to quote Ref.~\onlinecite{yl09}, `the appearance of $(2 \times 2)$
"clusters" that help to sustain the [$\,$collinear$\,$] order at
low temperatures even when the external field is slightly bigger
than $4$'. This sort of cluster is not taken into account
in our column-magnetization calculations, see Eq.~(\ref{eq:m_comb}).
We also know that the general effect of neglecting relevant contributions to
the magnetization is to increase the apparent value of $\beta/\nu$;
for instance, if $M_{\rm u}$ is discarded in Eq.~(\ref{eq:m_comb}),
the estimate of $\beta/\nu$ at $T=2.0820$, $H=0$ goes from $0.120(2)$
to $0.135(2)$. According to this interpretation, 
for $H \simeq 2.5$ or thereabouts, $(2 \times 2)$
configurations which are locally energetically favorable start 
contributing to ordering in the $N \to \infty$ limit, but are not captured
in the scheme of Eq.~(\ref{eq:m_comb}). With decreasing $T$ and increasing $H$,
the effect of such configurations becomes more relevant, providing a
mechanism through which the apparent exponent increases, although the 
real one, we conjecture, possibly increases a little but
stays slightly below $1/8$. This would be in line with the behavior of $\eta$, 
depicted in part (a) of Figure~\ref{fig:eta_cc}. 

Calculation of the thermal exponent $y_T$ via Eq.~(\ref{eq:y_prg}) in the
reentrant region gave negative values, an artifact already noticed
in Refs.~\onlinecite{kkg83,akg84}. However, evaluation of critical finite-size
susceptibilities at $T=0.35$ gives $\gamma/\nu$ in the range $1.71-1.80$,
depending on the details of corrections to scaling assumed for data fitting.
Although lacking in accuracy, this range of values is broadly consistent
both with the corresponding $\eta$ estimate, and with the hypothesis that
critical behavior obeys weak universality all along the collinear-paramagnetic
critical line.

\begin{figure}
{\centering \resizebox*{3.3in}{!}{\includegraphics*{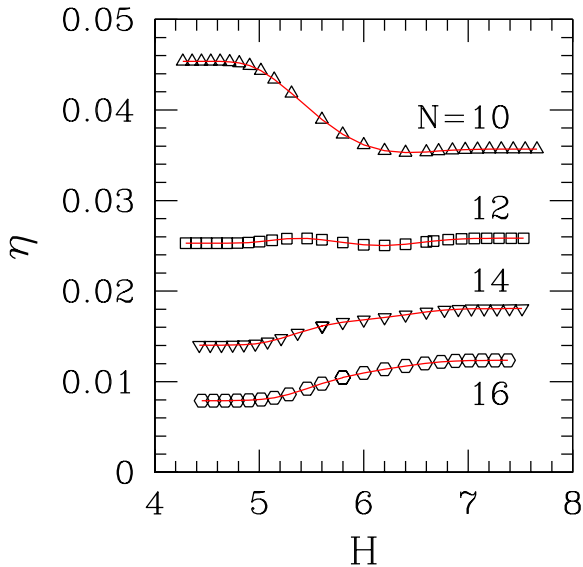}}}
\caption{(Color online) Decay-of-correlations exponent $\eta$,
evaluated via Eq.~(\protect{\ref{eq:eta_prg}}), along 
approximate row-shifted $(2 \times 2)$-paramagnetic transition lines,
obtained by solving Eq.~(\protect{\ref{eq:prg}}) for $N$ as indicated, for
high fields $4 \leq H \leq 8$.
} 
\label{fig:eta_hf}
\end{figure}

Turning now to the high-field part of the phase diagram, we 
illustrate in Figure~\ref{fig:eta_hf} our results for $\eta$ along
the approximate critical lines, the latter obtained by solving 
Eq.~(\ref{eq:prg}) for 
$N=10-16$. The curve for $N=12$ corresponds to that shown in Figure 2
of Ref.~\onlinecite{akg84}. 
By comparing the evolution of $\eta$ along the approximate critical lines
with the evolution of the lines themselves against increasing $N$,
shown in Figure~\ref{fig:hfpd}, one anticipates that calculating $\eta$
on the extrapolated phase boundary will give results very close to zero,
or even slightly negative. Indeed, this was what we found. It would appear
that this is at least partly because our extrapolation procedures
underestimate the extent of the row-shifted $(2 \times 2)$ phase.
Evaluation of  $c$ along the extrapolated critical line
also gave physically inconsistent results. 

Even though we are not able to produce numerically accurate estimates of $\eta$ 
for the high-field part of the phase diagram, the gist of the results shown in
Figure~\ref{fig:eta_hf} is that this must be below $0.01$, and possibly
even zero. We return to this point in the next Section.

\section{Discussion and Conclusions} 
\label{sec:conc}
For the model described by Eq.~(\ref{eq:def}), with $J_2=1$,
we have established a physical picture for the collinear-paramagnetic
phase boundary, which is consistent with continuously-varying
exponents along the critical line. Together with various
pieces of numerical evidence, collected at selected points, 
overall support for this is given by the conformal anomaly results depicted 
in part (b) of Figure~\ref{fig:eta_cc}.

There is also clear evidence that such continuously-varying exponents
satisfy, at least approximately, a weak-universality 
scenario~\cite{s74}. However, as shown for the exponent $\eta$ in part (a) 
of Figure~\ref{fig:eta_cc}, our results indicate small but consistent deviations
from the corresponding Ising values. 

Furthermore, such deviations are internally consistent, in the sense that
both $\eta$ and $\beta/\nu$ take on values lower than the Ising ones,
while $\gamma/\nu$ is always found to be higher than the Ising result. 
For $\beta/\nu$, the apparent reversal of this trend found for $H \gtrsim 2.5$
has been explained in Section~\ref{sec:h>0}, as a likely effect
of the same sort of locally stable $(2 \times 2)$ configurations which, 
at lower $T$ and higher $H$, become significant enough to induce reentrant 
behavior.

Notwithstanding the compensation just referred to,
our estimates of $(\gamma/\nu)+\eta$ in general
exceed $2$, though never by more than $2-3$ times the respective (combined) error bar.
However, it must be recalled that here $\gamma/\nu$ is essentially one order 
of magnitude larger than $\eta$,
and both quantities have similar relative 
uncertainties, thus the calculated combined uncertainty is practically only that
associated with $\gamma/\nu$. In such circumstances, the apparent violation
of a fundamental scaling relation reflects the fact that the relative
uncertainty in $\gamma/\nu$ was estimated as $\approx 2$ parts in $10^3$.
Had this been doubled, all the basic conclusions from this work would still
stand, and the mismatch would be essentially lost within the revised error
bars.  

Returning to $\eta$ as displayed in part (a) of Figure~\ref{fig:eta_cc},
the small but consistent shift between the high- and low-$T$ approximately
constant values [$\,$respectively, $0.2342(3)$ and $0.2476(3)\,$] indicates
a crossover between two distinct weak-universality classes. Such small
variations could probably be accounted for in the context of compactified
boson theory~\cite{dif97,k88}, in which continuously-varying critical indices
are put in direct correspondence with the (also continuously-varying)
radius $R$ associated with the underlying field theory.         

In general, both for $H=0$ and $H \neq 0$ (the latter, 
along the collinear-paramagnetic critical line) our results for the location of
critical points, and exponents such as $y_T$, $\gamma/\nu$, and $\beta/\nu$,
are mostly compatible, within error bars, with estimates available in the
literature. On the other hand, for the row-shifted ($2 \times 2)$-paramagnetic
phase boundary at high fields, we have found a discrepancy of
some $19\%$ between our estimate and that given in Ref.~\onlinecite{yl09},
for the highest transition temperature at $H=6$. Even allowing for the
(rather plausible) likelihood that our extrapolation procedures underestimate
the extent of the ordered phase, for PRG with 
the largest size available ($N=16$) one has $T_c(H=6)=0.651$, already 
$11\%$ below $T_c(H=6)=0.7293(7)$ quoted in Ref.~\onlinecite{yl09}.
At this point, such discrepancy remains unexplained.

We have not succeeded in gathering as much information
regarding critical properties of the row-shifted ($2 \times 2)$-paramagnetic
boundary line, as we did for its collinear-paramagnetic counterpart. 
However, the behavior of 
$\eta$ illustrated in Figure~\ref{fig:eta_hf} reminds one of the
low-temperature behavior of the two-dimensional $XY$ model. Indeed,
with $T_{KT}$ being the upper limit of the Kosterliz-Thouless
critical phase, the exponent $\eta$ of the $XY$ model 
grows smoothly and monotonically from
$\eta=0$ at $T=0$ to $1/4$ at $T_{KT}$ (with most of the
increase confined to higher $T$: at $T=0.5\,T_{KT}$, $
\eta \approx 0.05$)~\cite{bsp02}. So the very low values of $\eta$
found in the present case may, or may not, indicate the presence of
incipient $XY$-like behavior along at least part of the high-field
critical line.

\begin{acknowledgments}
The author thanks R. B. Stinchcombe, J. T. Chalker, and Fabian Essler 
for helpful discussions; thanks are due also to 
the Rudolf Peierls Centre for Theoretical Physics, Oxford, for the hospitality, 
and CAPES for funding the author's visit. 
The research of S.L.A.d.Q. is financed 
by the Brazilian agencies CAPES (Grant No. 0940-10-0),  
CNPq  (Grant No. 302924/2009-4), and FAPERJ (Grant No. E-26/101.572/2010).
\end{acknowledgments}


\begin{thebibliography}{99}
\bibitem{fm1} G. Misguich and C. Lhuillier, in
{\it Frustrated Spin Systems}, edited by H. T. Diep
(World Scientific, Singapore, 2005).
\bibitem{fm2} J. T. Chalker, in {\it Introduction to Frustrated
Magnetism: Materials, Experiment, Theory}, edited by C. Lacroix, P. Mendels, 
and F. Mila, Springer Series in Solid-State Sciences vol. 164
(Springer, Berlin, 2011).
\bibitem{yl09} Junqi Yin and D. P. Landau, \pre {\bf 80}, 051117 (2009).
\bibitem{kalz08} A. Kalz, A. Honecker, S. Fuchs, and T. Pruschke,
Eur. Phys. J. B {\bf 65}, 533 (2008).
\bibitem{fs2}
M. P. Nightingale, in {\it Finite Size Scaling and Numerical
Simulations of Statistical Systems}, edited by V. Privman (World Scientific,
Singapore, 1990).
\bibitem{barber} M. N. Barber, in {\it Phase Transitions and
Critical Phenomena}, edited by C. Domb and J. L. Lebowitz (Academic,
New York, 1983), Vol. 8.
\bibitem{cardy} J. L. Cardy, in {\it Phase Transitions and
Critical Phenomena}, edited by C. Domb and J. L. Lebowitz (Academic,
New York, 1987), Vol. 11.
\bibitem{bcn86} H. W. J. Bl\"ote, J. L. Cardy, and M. P. Nightingale,
\prl {\bf 56}, 742 (1986).
\bibitem{cardy84} J. L. Cardy, J. Phys. A {\bf 17}, L385 (1984).
\bibitem{kkg83} K. Kaski, W. Kinzel, and J. D. Gunton, \prb {\bf 27},
6777 (1983). 
\bibitem{akg84} J. Amar, K. Kaski, and J. D. Gunton, \prb {\bf 29},
1462 (1984). 
\bibitem{nb98} M. P. Nightingale and  H. W. J. Bl\"ote, Physica A {\bf 251},
211 (1998).
\bibitem{dds82} B. Derrida and L. de Seze, J. Phys. (Paris) {\bf 43}, 475 (1982).
\bibitem{nb82} M. P. Nightingale and  H. W. J. Bl\"ote,  J. Phys. A {\bf 15}, 
L33 (1983).
\bibitem{pf83} V. Privman and M.E. Fisher, J. Phys. A {\bf 16}, 
L295 (1983).
\bibitem{bg85} T. W. Burkhardt and I. Guim, J. Phys. A {\bf 18}, L25 (1985).
\bibitem{bg85b} T. W. Burkhardt and I. Guim, J. Phys. A {\bf 18}, L33 (1985).
\bibitem{bn82} H. W. J. Bl\"ote and M. P. Nightingale,  Physica A {\bf 112}, 
405 (1982).
\bibitem{bn85} H. W. J. Bl\"ote and M. P. Nightingale, Physica A {\bf 134},
274 (1985).
\bibitem{nb83} M. P. Nightingale and  H. W. J. Bl\"ote, J. Phys. A {\bf 16}, 
L657 (1983).
\bibitem{hamer82} C. J. Hamer, J. Phys. A {\bf 15}, 
L675 (1982).
\bibitem{cardy87}  J. L. Cardy, J. Phys. A {\bf 20}, L891 (1987).
\bibitem{dif97} P. di Francesco, P. Mathieu, and D. S\'en\'echal,
{\it Conformal Field Theory} (Springer-Verlag, New York, 1997). 
\bibitem{kalz11} A. Kalz, A. Honecker, and M. Moliner,
arXiv:1105.4836 (2011).
\bibitem{has08} M. Hasenbusch, F. Parisen Toldin, A. Pelissetto, and E. Vicari,
\pre {\bf 77}, 051115 (2008).
\bibitem{dq09} S. L. A. de Queiroz, \prb {\bf 79}, 174408 (2009).
\bibitem{mkt06} A. Malakis, P. Kalozoumis, and N. Tyraskis, Eur. Phys. J. B
{\bf 50}, 63 (2006).
\bibitem{mk07} J. L. Monroe and S.-Y. Kim, \pre {\bf 76}, 021123 (2007).
\bibitem{kim10} S.-Y. Kim, \pre {\bf 81}, 031120 (2010).
\bibitem{s74} M. Suzuki, Prog. Theor. Phys. {\bf 51}, 1992 (1974).
\bibitem{bl80} K. Binder and D. P. Landau, \prb {\bf 21}, 1941 (1980).
\bibitem{k87} M. Kaufman, \prb {\bf 36}, 3697 (1987).
\bibitem{dq09b} S. L. A. de Queiroz, \pre {\bf 80}, 041125 (2009).
\bibitem{slotte} P. A. Slotte, J. Phys. C {\bf 16}, 2935 (1983).
\bibitem{k88} M. Karowski, Nucl. Phys. {\bf B300}, 473 (1988).
\bibitem{bsp02} B. Berche, A. I. F. Sanchez, and R. Paredes,
Europhys. Lett. {\bf 60}, 539 (2002).


\end{thebibliography}
\end{document}